\documentclass[journal,12pt,onecolumn,draftclsnofoot,]{IEEEtran}
\usepackage{cite}

\ifCLASSINFOpdf
  \usepackage[pdftex]{graphicx}
\else
  \usepackage[dvips]{graphicx}
\fi
\usepackage{amsmath,amssymb}
\usepackage{array}
\usepackage{multirow}
\usepackage{hhline}
\usepackage[draft,ulem=normalem]{changes}

\begin{document}
\newif\ifdraft
\draftfalse 
%
\title{On the Safety of Machine Learning: \\ {\huge Cyber-Physical Systems, Decision Sciences, and Data Products}}

\author{\IEEEauthorblockN{Kush R. Varshney (Corresponding Author)\\}
\IEEEauthorblockA{Data Science Theory and Algorithms\\
IBM Thomas J. Watson Research Center\\
Yorktown Heights, New York 10598\\
Email: krvarshn@us.ibm.com\\}
\and
\IEEEauthorblockN{Homa Alemzadeh\\}
\IEEEauthorblockA{Electrical and Computer Engineering\\
University of Virginia\\
Charlottesville, Virginia 22904\\
Email: alemzadeh@virginia.edu}}


%


\maketitle

\begin{abstract}
Machine learning algorithms increasingly influence our decisions and interact with us in all parts of our daily lives. Therefore, just as we consider the safety of power plants, highways, and a variety of other engineered socio-technical systems, we must also take into account the safety of systems involving machine learning. Heretofore, the definition of safety has not been formalized in a machine learning context. In this paper, we do so by defining machine learning safety in terms of risk, epistemic uncertainty, and the harm incurred by unwanted outcomes. We then use this definition to examine safety in all sorts of applications in cyber-physical systems, decision sciences, and data products. We find that the foundational principle of modern statistical machine learning, empirical risk minimization, is not always a sufficient objective. Finally, we discuss how four different categories of strategies for achieving safety in engineering, including inherently safe design, safety reserves, safe fail, and procedural safeguards can be mapped to a machine learning context. We then discuss example techniques that can be adopted in each category, such as considering interpretability and causality of predictive models, objective functions beyond expected prediction accuracy, human involvement for labeling difficult or rare examples, and user experience design of software and open data.
\end{abstract}


%
\IEEEpeerreviewmaketitle

\section{Introduction}
\label{sec:intro}

In recent years, machine learning algorithms have started influencing every part of our lives, including health and wellness, law and order, commerce, entertainment, finance, human capital management, communication, transportation, and philanthropy.  As the algorithms, the data on which they are trained, and the models they produce are getting more powerful and more ingrained in society, questions about \emph{safety} must be examined.  It may be argued that machine learning systems are simply tools, that they will soon have a general intelligence that surpasses human abilities, or something in-between. But from all perspectives, they are technological components of larger socio-technical systems that may have to be engineered with safety in mind  \cite{Conn2015}.

Safety is a commonly used term across engineering disciplines connoting the absence of failures or conditions that render a system dangerous \cite{Ferrell2010}. Safety is a notion that is domain-specific, cf.\ safe food and water, safe vehicles and highways, safe medical treatments, safe toys, safe neighborhoods, and safe industrial plants.  Each of these domains has specific design principles and regulations that are applicable only to them.  

There are some loose notions of safety for machine learning, but they are primarily of the ``I know it when I see it'' variety or are very application-specific; to the best of our knowledge \cite{Varshney2016}, there is no precise, non-application-specific, first-principles definition of safety for machine learning.  The main contribution of this paper is to provide exactly such a definition.  To do so, we build upon a universal domain-agnostic definition of safety in the engineering literature \cite{MollerH2008,Moller2012}.  

In \cite{MollerH2008,Moller2012} and numerous references therein, Moeller et al. propose a decision-theoretic definition of safety that applies to a broad set of domains and systems. They define safety to be the reduction or minimization of \emph{risk} and \emph{epistemic uncertainty} associated with unwanted outcomes that are severe enough to be seen as \emph{harmful}. The key points in this definition are: i) the cost of unwanted outcomes has to be sufficiently high in some human sense for events to be harmful, and ii) safety involves reducing both the probability of expected harms and the possibility of unexpected harms. 

We define safety in machine learning in the same way, as the minimization of both risk and uncertainty of harms, and devote Section~\ref{sec:ml} to fleshing out the details of this definition. As such, formulations of machine learning for achieving safety that we describe in Section~\ref{sec:ml:strat} must have both risk and uncertainty minimization in their objective functions either explicitly, implicitly via constraints, or through socio-technical components beyond the core machine learning algorithm.  The harmful cost regime is the part of the space that requires the dual objectives of risk and uncertainty minimization; the non-harmful cost regime does not require the uncertainty minimization objective.

As background before getting to those sections, we briefly describe harms, risk, and uncertainty without specialization to machine learning.  A system yields an outcome based on its state and the inputs it receives. An outcome event may be desired or undesired. Single events and sets of events have associated costs that can be measured and quantified by society. For example, a numeric level of morbidity can be the cost of an outcome.  An undesired outcome is only a harm if its cost exceeds some threshold. Unwanted events of small severity are not counted as safety issues. Risk is the expected value of the cost.  Epistemic uncertainty results from the lack of knowledge that could be obtained in principle, but may be practically intractable to gather \cite{SengeBDHHDH2014}. Harmful outcomes often occur in regimes and operating conditions that are unexpected or undetermined. With risk, we do not know what the outcome will be, but its distribution is known, and we can calculate the expectation of its cost. With uncertainty, we still do not know what the outcome will be, but in contrast to risk, its probability distribution is also unknown (or only partially known). Some decision theorists argue that all uncertainty can be captured probabilistically, but we maintain the distinction between risk and uncertainty \cite{Moller2012}.

The first contribution of this work is to critically examine the foundational statistical machine learning principles of empirical risk minimization and structural risk minimization \cite{Vapnik1992} from the perspective of safety.  We discuss how they do not deal with epistemic uncertainty.  Further, these principles rely on arguments involving average losses and laws of large numbers, which may not necessarily be fully applicable when considering safety.  Moreover, the loss functions involved in these principles are abstract measures of distance between true and predicted values rather than application-specific quantities measuring the possibility of outcomes such as loss of life or loss of quality of life that can be judged harmful or not \cite{Wagstaff2012}.  

A discussion of safety would be incomplete without a discussion of strategies to increase the safety of socio-technical systems with machine learning components.  Four categories of approaches have been identified for promoting safety in general \cite{MollerH2008}: inherently safe design, safety reserves, safe fail, and procedural safeguards.  As a second contribution, we discuss these approaches specifically for machine learning algorithms and especially to mitigate epistemic uncertainty.  Through this contribution, we can recommend strategies to engineer safer machine learning methods and set an agenda for further machine learning safety research.

The third contribution of this paper is examining the definition of and strategies for safety in specific machine learning applications. Today, machine learning technologies are used in a variety of settings, including cyber-physical systems, decision sciences, and data products.  By cyber-physical systems, we mean engineered systems that integrate computational algorithms and physical components, e.g.\ surgical robots, self-driving cars, and the smart grid \cite{Alemzadeh2016}. By decision sciences, we mean the use of algorithms to aid people in making important decisions and informing strategy, e.g.\ prison parole, medical treatment, and loan approval \cite{StanleyT2016}. By data products, we mean the use of algorithms to automate informational products, e.g.\ web advertising placement, media recommendation, and spam filtering \cite{StanleyT2016}. These settings vary widely in terms of their interaction with people, the scale of data, the time scale of operation and consequence, and the cost magnitude of consequences.  A further contribution is a discussion on how to even understand and quantify the desirability and undesirability of outcomes along with their costs.  To complement simply eliciting such knowledge directly from people \cite{OlteanuTV2017}, we suggest a data-driven approach for characterizing harms that are particularly relevant for cyber-physical systems with large state spaces of outcomes.

Overall, the purpose of this paper is to introduce a common language and framework for understanding, evaluating, and designing machine learning systems that involve society and technology.  
Our goal is to set forth a fundamental organizing and unifying principle that carries through to abstract theoretical formulations of machine learning as well as to concrete real-world applications of machine learning.  Thus it provides practitioners working at any level of abstraction a principled way to reason about the space of socio-technical solutions.

The remainder of the paper is organized in the following manner. In Section~\ref{sec:ml}, after introducing the standard notation and concept of statistical machine learning, we discuss what harm, risk, and epistemic uncertainty mean for machine learning. In Section~\ref{sec:ml:strat},  we discuss specific strategies for achieving safety in machine learning. Section~\ref{sec:example} dives into example applications in cyber-physical systems, decision sciences, and data products. Section~\ref{sec:conclusion} concludes the paper.

\section{Safety in Machine Learning}
\label{sec:ml}

In this section, after briefly introducing statistical machine learning notation, we examine how machine learning applications fit with the conception of safety given above.

\subsection{Notation}

In what follows, we use standard notation to describe concepts from empirical risk minimization \cite{Vapnik1992}. Given joint random variables $X \in \mathcal{X}$ (features) and $Y \in \mathcal{Y}$ (labels) with probability density function $f_{X,Y}(x,y)$, a function mapping $h \in \mathcal{H}: \mathcal{X} \rightarrow \mathcal{Y}$, and a loss function $L : \mathcal{Y} \times \mathcal{Y} \rightarrow \mathbb{R}$, the risk $R(h)$ is defined as the expected value of loss: $$\mathbb{E}[L(h(X),Y)]= \int_\mathcal{X}\int_\mathcal{Y}L(h(x),y)f_{X,Y}(x,y)dydx.$$

The loss function $L$ typically measures the discrepancy between the value predicted for $y$ using $h(x)$ and $y$ itself, for example $(h(x)-y)^2$ in regression problems. We would like to learn the function $h$ that minimizes the risk.

In the machine learning context, we do not have access to the probability $f_{X,Y}$, but rather to a training set of samples drawn i.i.d.\ from the joint distribution $(X,Y)$: $\{(x_1,y_1),\ldots,(x_m,y_m)\}$ and the goal is to learn $h$ such that the empirical risk $R^{emp}_m(h)$ is minimized. The emprical risk is given by: $$R^{emp}_m(h) = \frac{1}{m}\sum_{i=1}^m L(h(x_i),y_i).$$

\subsection{Harmful Costs}
\label{sec:ml:costs}

Analyzing safety requires us first to examine whether immediate human costs of outcomes exceed some severity threshold to be harmful.  Unlike other domains mentioned in the introduction, such as safe industrial plants and safe toys, we have a great advantage when working with machine learning systems because the optimization formulation explicitly includes the loss function $L$. The domain of $L$ is $\mathcal{Y}\times\mathcal{Y}$ and the output is an abstract quantity representing prediction error.  In real-world applications, the value of the loss function may be endowed with some human cost and that human cost may imply a loss function that also includes $\mathcal{X}$ in the domain. Moreover, the cost may be severe enough to be harmful and thus a safety issue in some parts of the domain and not in others.  

In many decision science applications, undesired outcomes are truly harmful in a human sense and their effect is felt in near-real time.  They are safety issues. Moreover, the space of outcomes is often binary or of small cardinality and it is often self-evident which outcomes are undesired.  However,  loss functions are not always monotonic in the correctness of predictions and depend on whose perspective is in the objective. The space of outcomes for the machine learning components of typical cyber-physical systems applications is so vast that it is near-impossible to enumerate all of the outcomes, let alone elicit costs for them.  Nevertheless, it is clear that outcomes leading to accidents have high human cost in real time and require the consideration of safety.  In order to get more nuanced characterizations of the cost severity of outcomes, a data-driven approach is prudent \cite{AlemPLOS2016}. The quality of service implications of unwanted outcomes in data product applications are not typically safety hazards because they do not have an immediate severe human cost. Undesired outcomes may only hypothetically lead to human consequences. In practice, often the  acceptable levels of safety and accident rates are defined by the society and the application domain. For example, the difference in acceptable accident rates and costs in motor vehicles (hundreds of thousands of fatalities per year) versus commercial aircraft (tens of fatalities per year) shows the subjectivity of the public's acceptance of safety \cite{Knight2012}. 

\subsection{Risk and Epistemic Uncertainty}

The risk minimization approach to machine learning has many strengths, which is evident by its successful application in various domains. We benefit from this explicit optimization formulation in the machine learning domain by automatically reducing the probability of harms, which is not always the case in other domains.  However, this standard formulation does not capture the issues related to the uncertainty that are also relevant for safety.  

First, although it is assumed that the training samples $\{(x_1,y_1),\ldots,(x_m,y_m)\}$ are drawn from the true underlying probability distribution of $(X,Y)$, that may not always be the case.  Further, it may be that the distribution the samples actually come from cannot be known, precluding the use of covariate shift \cite{shimodaira2000} and domain adaptation techniques \cite{daume2006}.  This is one form of epistemic uncertainty that is quite relevant to safety because training on a dataset from a different distribution can cause much harm.

Also, it may be that the training samples do come from the true, but unknown, underlying distribution, but are absent from large parts of the $\mathcal{X}\times\mathcal{Y}$ space due to small probability density there.  Here the learned function $h$ will be completely dependent on an inductive bias encoded through $\mathcal{H}$ rather than the uncertain true distribution, which could introduce a safety hazard.

The statistical learning theory analysis utilizes laws of large numbers to study the effect of finite training data and the convergence of $R^{emp}_m(h)$ to $R(h)$. However, when considering safety, we should also be cognizant that in practice, a machine learning system only encounters a finite number of test samples and the actual operational risk is an empirical quantity on the test set.  Thus the operational risk may be much larger than the actual risk for small cardinality test sets, even if $h$ is risk-optimal.  This uncertainty caused by the instantiation of the test set can have large safety implications on individual test samples.

Applications performed at scales with large training sets, large testing sets, and the ability to explore the feature space have little epistemic uncertainty, whereas in other applications it is more often than not the case that there is uncertainty about the training samples being representative of the testing samples and that only a few predictions are made.  Moreover, in applications such as cyber-physical systems, very large outcome spaces prevent even mild coverage of the space through training samples. 

\section{Strategies for Achieving Safety}
\label{sec:ml:strat}

As discussed, safety and strategies for achieving it are often investigated on an application-by-application basis.  For example, setting the minimum thickness of vessels and removing flammable materials from a chemical plant are ways of achieving safety.  By analyzing such strategies across domains, \cite{MollerH2008} has identified four main categories of approaches to achieve safety. 

First, inherently safe design is the exclusion of a potential hazard from the system (instead of controlling the hazard).  For example, excluding hydrogen from the buoyant material of a dirigible airship makes it safe.  (Another possible safety measure would be to introduce apparatus to prevent the hydrogen from igniting.)  

A second strategy for achieving safety is through multiplicative or additive reserves, known as safety factors and safety margins, respectively.  In mechanical systems, a safety factor is a ratio between the maximal load that does not lead to failure and the load for which the system was designed.  Similarly, the safety margin is the difference between the two.  

The third general category of safety measures is `safe fail,' which implies that a system remains safe when it fails in its intended operation.  Examples are electrical fuses, so-called dead man's switches on trains, and safety valves on boilers.  

Finally, the fourth strategy for achieving safety is given the name procedural safeguards.  This strategy includes measures beyond ones designed into the core functionality of the system, such as audits, training, posted warnings, and so on.  

In this section, we discuss each of these strategies with specific approaches that extend machine learning formulations beyond risk minimization for safety.

\subsubsection{Inherently Safe Design}
\label{sec:ml:strat:inherent}

In the machine learning context, we would like robustness against the uncertainty of the training set not being sampled from the test distribution. The training set may have various biases that are unknown to the user and that will not be present during the test phase or may contain patterns that are undesired and might lead to harmful outcomes. Modern techniques such as extreme gradient boosting and deep neural networks may exploit these biases and achieve high accuracy, but they may fail in making safe predictions due to unknown shifts in the data domain or inferring incorrect patterns or harmful rules \cite{CaruanaLGKSE2015}.  

These models are so complex that it is very difficult to understand how they will react to such shifts and whether they will produce harmful outcomes as a result.  Two related ways to introduce inherently safe design are by insisting on models that can be interpreted by people and by excluding features that are not causally related to the outcome \cite{Freitas2013,Rudin2014,AtheyI2015,Welling2015}.  By examining interpretable models, features or functions capturing quirks in the data can be noted and excluded, thereby avoiding related harm.  Similarly, by carefully selecting variables that are causally related to the outcome, phenomena that are not a part of the true `physics' of the system can be excluded, and associated harm be avoided. We note that post hoc interpretation and repair of complex uninterpretable models, appealing for other reasons, does not assure safety via inherently safe design because the interpretation is not the decision rule that is actually used in making predictions.

Neither interpretability nor causality of models is properly captured within the standard risk minimization formulation of machine learning.  Extra regularization or constraints on $\mathcal{H}$, beyond those implied by structural risk minimization, are needed to learn inherently safe models. That might lead to performance loss in accuracy when measured through standard metrics such as training and testing data probability distributions, but the safety will be enhanced by reduction in epistemic uncertainty and undesired bias. Both interpretability and causality may be incorporated into a single learned model, e.g.~\cite{WangR2015b}, and causality may be used to induce interpretability, e.g.~\cite{ChakarovNRSV2016}. In applications with very large outcome spaces such as those employing reinforcement learning, it is shown that appropriate aggregation of states in outcome policies can lead to interpretable models\cite{PetrikL2016}.

\subsubsection{Safety Reserves}
\label{sec:ml:strat:reserves}

In machine learning formulations, the uncertainty in the matching of training and test data distributions or in the instantiation of the test set can be parameterized with the symbol $\theta$.  Let $R^*(\theta)$ be the risk of the risk-optimal model if the $\theta$ were known. Along the same lines as safety factors and safety margins, robust formulations find $h$ while constraining or minimizing $\max_\theta \frac{R(h,\theta)}{R^*(\theta)}$ or $\max_\theta \left(R(h,\theta) - R^*(\theta)\right)$.  Such formulations can capture uncertainty in the class priors and uncertainty resulting from label noise in classification problems \cite{ProvostF2001,DavenportBS2010}. They can also capture the uncertainty of which part of the $\mathcal{X}$ space the actual small set of test samples comes from.

A different sort of safety factor comes about when considering fairness and equitability. In certain prediction problems, the risk of harm for members of protected groups should not be much worse (up to a multiplicative factor) than the risk of harm for others \cite{HajianD2013,FeldmanFMSV2015,BarocasS2016}. We can partition the feature space $\mathcal{X}$ into the sets  $\mathcal{X}_u, \mathcal{X}_p \subset \mathcal{X}$, respectively, corresponding to the unprotected and protected groups, indicated by features such as race and gender. Then using a rule such as the 80\% (or four-fifths) rule advocated in the study of disparate impact \cite{EEOC1979}, we can constraint the relative risk of harm for the protected versus unprotected group to a maximum value such as $5/4$: $$\frac{\int_{\mathcal{X}_p}\int_\mathcal{Y}L(x,h(x),y)f_{X,Y}(x,y)dydx}{\int_{\mathcal{X}_u}\int_\mathcal{Y}L(x,h(x),y)f_{X,Y}(x,y)dydx}\le\frac{5}{4}.$$ Under such a constraint, we ensure that the outcome of prediction for protected groups is not much more harmful than for unprotected groups.

\subsubsection{Safe Fail}
\label{sec:ml:strat:fail}

A technique used in machine learning when predictions cannot be given confidently is the reject option \cite{VarshneyPMCH2013}: the model reports that it cannot reliably give a prediction and does not attempt to do so, thereby failing safely.  When the model selects the reject option, typically a human operator intervenes, examines the test sample, and provides a manual prediction.

In classification problems, models are reported to be least confident near the decision boundary.  However, by doing so, there is an implicit assumption that distance from the decision boundary is inversely related to confidence.  This is reasonable in parts of $\mathcal{X}$ with high probability density and large numbers of training samples because the decision boundary is located where there is a large overlap in likelihood functions.  However parts of $\mathcal{X}$ with low density may not contain any training samples at all and the decision boundary may be completely based on an inductive bias, thereby containing much epistemic uncertainty.  In these parts of the space, distance from the decision boundary is fairly meaningless and the typical trigger for the reject option should be avoided \cite{AttenbergIP2015}.  For a rare combination of features in a test sample \cite{Weiss2004}, a safe fail mechanism is to always go for manual examination.

Both of these manual intervention options are suitable for applications with sufficiently long time scales.  When working on the scale of milliseconds, only options similar to dead man's switches that stop operations in a reasonable manner are applicable.

\subsubsection{Procedural Safeguards}
\label{sec:ml:strat:procedure}

In addition to general procedural safeguards that carry over from other domains, two directions in machine learning that can be used for increasing safety within this category are user experience design and openness.  

In decision science applications especially, non-specialists are often the operators of machine learning systems.  Defining the training data set and setting up evaluation procedures, among other things, have certain subtleties that can cause harm during operation if done incorrectly.  User experience design can be used to guide and warn novice and experienced practitioners to set up machine learning systems properly and thereby increase safety.

These days most modern machine learning algorithms are open source, which allows for the possibility of the public audit.  Safety hazards and potential harms can be discovered through examination of source code.  However, open source software is not sufficient, because the behavior of machine learning systems is driven by data as much as it is driven by software implementations of algorithms.  Open data refers to data that can be freely used, reused, and redistributed by anyone. Opening data is a procedural safeguard for increasing safety that is increasingly being adopted by the community \cite{SahuguetKPS2014,Shaw2015,KapoorMSV2015}.  

\section{Example Applications}
\label{sec:example}

In this section, we further detail safety in machine learning systems by providing examples from cyber-physical systems, decision sciences, and data products.

\subsection{Cyber-Physical Systems}
With advances in computing, networking, and sensing technologies, cyber-physical systems have been deployed in various safety-critical settings such as aerospace, energy, transportation, and healthcare. The increasing complexity and  connectivity of these systems, the tight coupling between their cyber and physical components, and the inevitable involvement of human operators in their supervision and control has introduced significant challenges in ensuring system reliability and safety while maintaining the expected performance. Cyber-physical systems continuously interact with the physical world and human operators in real-time. In order to adapt to the constantly changing and uncertain environment, they need to take into account not only the current application but also the operator's preferences, intent, and past behavior \cite{Schirner2013}. 

Autonomous machine learning and artificial intelligence techniques have been applied to several decision-making and control problems in cyber-physical systems. Here we discuss two examples where unexpected harmful events with epistemic uncertainty might impact human lives in real-time.

\subsubsection{Surgical Robots}

Robotically-assisted surgical systems are a typical example of human-in-the-loop cyber-physical systems. Surgical robots consist of a teleoperation console operated by a surgeon, an embedded system hosting the automated robot control, and the physical robotic actuators and sensors. The robot control system receives the surgeon's commands issued using the teleoperation console and translates the surgeon's hand, wrist, and finger movements into precisely engineered movements of miniaturized surgical instruments inside patient's body. Recent research shows an increasing interest in the use of machine learning algorithms for modeling surgical skills, workflow, and environment and integration of this knowledge into control and automation of surgical robots \cite{Kassahun2016}. Machine learning techniques have been used for detection and classification of surgical motions for automated surgical skill evaluation \cite{Lin2005,Lin2006, Reiley2010} and automating portions of repetitive and time-consuming surgical tasks (e.g., knot-tying, suturing).\cite{Reiley2010, Shademan2016}. 

In autonomous robotic surgery, a machine learning enabled surgical robot continuously estimates the state of the environment (e.g., length or thickness of soft tissues under surgery) based on the measurements from sensors (e.g., image data or force signals) and generates a plan for executing actions (e.g., moving the robotic instruments along a trajectory). The mapping function from the perception of environment to the robotic actions is considered as a surgical skill which the robot learns, through evaluation of its own actions or from observing the actions of expert surgeons. The quality of the learned surgical skills can be assessed using cost functions that are either automatically learned or are manually defined by surgeons \cite{Kassahun2016}.  

Given the uncertainty and large variability in the operator actions and behavior, organ/tissue movements and dynamics, and possibility of incidental failures in the robotic system and instruments, predicting all possible system states and outcomes and assessing their associated costs is very challenging. As mentioned in Section \ref{sec:ml:costs}, due to the very large outcome space, it is not straightforward to elicit costs of all different outcomes and characterize which tasks or actions are costly enough to represent safety issues. For example, there have been ongoing reports of safety incidents during use of surgical robots that negatively impact patients by causing procedure interruptions or minor injuries. These incidents happen despite existing safe fail mechanisms included in the system and often result from a combination of different causal factors and unexpected conditions, including malfunctions of surgical instruments, actions taken by the surgeon, and the patient's medical history\cite{AlemPLOS2016}. 

There are also practical limitations in learning optimal and safe surgical trajectories and workflows due to epistemic uncertainty in such environments. The training data often consists of samples collected from a select set of surgical tasks (e.g., elementary suturing gestures) performed by well-trained surgeons, which might not represent the variety of actions and tasks performed during a real procedure. Previous work shows that surgeon's expertise level, surgery type, and medical history have a significant impact on the possibility of complications and errors occurring during surgery.  Further, automated algorithms should be able to cope with uncertainty and unpredictable events and guarantee patient safety just as expert surgeons do in such scenarios \cite{Kassahun2016}. 

One solution for dealing with these uncertainties is to assess the robustness of the system in the presence of unwanted and rare hazardous events (e.g., failures in control system, noisy sensor measurements, or incorrect commands sent by novice operators) by simulating such events in virtual environments \cite{AlemSAFECOMP2015} and quantifying the possibility of making safe decisions by the learning algorithm. This approach is an example of procedural safeguards (Section \ref{sec:ml:strat:procedure}). Such a simulated assessment also serves to highlight the situations requiring safe fail strategies, such as converting the procedure to non-robotic techniques, rescheduling it to a later time, or restarting the system, that can refine the system. The costs of unwanted outcomes and safe fail strategies to cope with them can also be characterized based on past data. For example, we mined the FDA's Manufacturer and User Facility Device Experience (MAUDE) database, a large database containing 14 years worth of adverse events, to obtain such characterizations on the causes and severity of safety incidents and recovery actions taken by the surgical team. Such analysis helps focus development of machine learning algorithms containing safety strategies on regimes with harmful outcomes and avoid concern for safety strategies in regimes with non-harmful outcomes. 

Another solution currently adopted in practice is through supervisory control of automated surgical tasks instead of fully autonomous surgery. For example, if the robot generates a geometrically optimized suture plan based on sensor data or surgeon input, it should still be tracked and updated in real time because of possible tissue motion and deformation during surgery \cite{Shademan2016}. This is an example of examining interpretable models to avoid possible harm (as discussed in Section \ref{sec:ml:strat:inherent}). An example of adopting safety reserves (Section \ref{sec:ml:strat:reserves}) in robotic surgery is robust optimization of preoperative planning to minimize the uncertainty at the task level while maximizing the dexterity \cite{AzimianNKP2015}.

\subsubsection{Self-Driving Cars}

Self-driving cars are autonomous cyber-physical systems capable of making intelligent navigation decisions in real-time without any human input. They combine a range of sensor data from laser range finders and radars with video and GPS data to generate a detailed 3D map of the environment and estimate their position. The control system of the car uses this information to determine the optimal path to the destination and sends the relevant commands to actuators that control the steering, braking, and throttle. Machine learning algorithms are used in the control system of self-driving cars to model, identify, and track the dynamic environment, including the road conditions and moving objects (e.g., other cars and pedestrians). 

Although automated driving systems are expected to eliminate human driver errors and reduce the possibility of crashes, there are several sources of uncertainty and failure that might lead to potential safety hazards in these systems. Unreliable or noisy sensor signals (e.g., GPS data or video signals in bad weather conditions), limitations of computer vision systems, and unexpected changes in the environment (e.g., unknown driving scenes or unexpected accidents on the road) can adversely affect the ability of control system in learning and understanding the environment and making safe decisions \cite{Robohub2014}. For example, a self-driving car (in auto-pilot mode) recently collided with a truck after failing to apply the brakes, leading to the death of the truck driver. This was the first known fatality in over 130 million miles of testing the automated driving system. The accident was caused under extremely rare circumstances of the high height of the truck, its white color under the bright sky, combined with the positioning of the cars across the road \cite{TeslaAccident2016}. 

The importance of epistemic uncertainty or "uncertainty on uncertainty" in these AI-assisted systems has been recently recognized, and there are ongoing research efforts towards quantifying the robustness of self-driving cars to events that are rare (e.g., distance to a bicycle running on an expected trajectory) or not present in the training data (e.g., unexpected trajectories of moving objects) \cite{UncertainOnUncertain}.  Systems that recognize such rare events trigger safe fail mechanisms.  

To the best of our knowledge, there is no self-driving car system with an inherently safe design that utilizes, e.g., interpretable models \cite{ZhuJ2017}.  Fail-safe mechanisms that upon detection of failures or less confident predictions, stop the autonomous control software and switch to a backup system or a degraded level of autonomy (e.g., full control by the driver) are considered for self-driving cars \cite{koopman2016}. 

\subsection{Decision Sciences}

In decision sciences applications, people are in the loop in a different way than in cyber-physical systems, but in the loop nonetheless.  Decisions are made about people and are made by people using machine learning-based tools for support.  Many emerging application domains are now shifting to data-driven decision making due to a greater capture of information digitally and the desire to be more scientific rather than relying on (fallible) gut instinct \cite{BrynjolfssonHK2011}.  These applications present many safety-related challenges.

\subsubsection{Predicting Voluntary Resignation}

We recently studied the problem of predicting which IBM employees will voluntarily resign from the company in the next six months based on human resources and compensation data, which required us to develop a classification algorithm to be placed within a larger decision-making system involving human decision makers \cite{SinghVWMGFE2012}.  There are several sources of epistemic uncertainty in this problem.  First, the way to construct a training set in the problem is to look at the historical set of employees and treat employees that voluntarily resigned as positive samples and employees still in the workforce as negative samples.  However, since the prediction problem is to predict resignation in the next six months, our set of negative samples will necessarily include employees who should be labeled positively because they will be resigning soon \cite{WeiV2015}.  

Another uncertainty is related to quirks or vagaries in the data that are predictive but will not generalize.  In this problem, a few predictive features related to stipulations in employees' contracts to remain with IBM for a fixed duration after their company was acquired, but such a pattern would not remain true going forward.  Another issue is unique feature vectors: if the data contains an employee in Australia who has gone 17 years without being promoted and no other similar employees, then there is huge uncertainty in that part of feature space, and inductive bias must be completely relied upon.

In the solution created for this problem, the inherently safe design principle of interpretability (Section \ref{sec:ml:strat:inherent}) was insisted upon and was what led to the discovery about the acquired company.  Specifically, C5.0 decision trees were used with the rule set option, and the project directly motivated the study of an optimization approach for learning classification rules \cite{MalioutovV2013}.  The reason for conducting the project was to take actions such as salary increases to retain employees at risk of resigning, and for this, the other inherently safe design principle of causality is important.  Rare samples such as the Australian employee led to the safe fail mechanism of manual inspection.  

\subsubsection{Loan Approval}

As another example in the decision sciences that we have studied, let us consider the decision to approve loans for solar panels given to the rural poor in India based on data in application forms \cite{GerardRSVKN2015}.  The epistemic uncertainty related to the training set not being representative of the true test distribution repeat here and can be addressed by similar safety strategies as discussed in the previous examples.  

Loan approval is an example illustrating that loss functions that are not always monotonic in the correctness of predictions and depend on perspective.  The applicant would like an approval decision regardless of their features indicating ability to repay, the lender would like approval only in cases in which applicant features indicate likely repayment, and society would like there to be fairness or equitability in the system so that protected groups, such as defined by gender and religion, are not discriminated against.  The lender perspective is consistent with the typical choice of the loss function, but the others are not.

An interesting additional issue, in this case, relates to the human cost function from society's perspective including $\mathcal{X}$.  One of the attributes available in the problem was the surname of the applicant; in this part of India, the surname is a strong indicator of religion and caste.  The use of this variable as a feature improved classification accuracy by a couple of percentage points, but resulted in worse fairness: the true cost in the problem from society's perspective.  Simply dropping the attribute as a feature does not ensure fairness because other features may be correlated, but a safety margin on the accuracy of the groups make the system fairer.

\subsection{Data Products}

With data products applications, the first question to consider is whether immediate costs are large enough for them to be considered safety issues.  One may argue that an algorithm showing biased or misguided advertisements or a spam filter not allowing an important email to pass could eventually lead to harm, e.g., by being shown an ad for a lower-paying job rather than a higher-paying one, a person may hypothetically end up with a lower quality of life at some point in the future.  Here the cost function does depend on $\mathcal{X}$ because misclassifying certain emails is more costly than others.  However, we do not view such a delayed and only hypothetical consequence as a safety issue.  

Moreover, in typical data products applications, one can use billions of data points as training, perform large-scale A/B testing, and evaluate average performance on millions or billions of clicks.  Therefore, uncertainty is not at the forefront, and neither are the safety strategies.  For example, the procedural safeguard of opening data is more common in decision science applications such as those sponsored or run by governments than in data products applications where the data is often the key value proposition.

\section{Conclusion}
\label{sec:conclusion}

Machine learning systems are already embedded in many functions of society.  The prognosis is for broad adoption to only increase across all areas of life.  With this prevailing trend, researchers, engineers, and ethicists have started discussing the topic of safety in machine learning.  In this paper, we contribute to this discussion starting from a very basic definition of safety in terms of harm, risk, and uncertainty and building upon it in the machine learning context.  We identify that the minimization of epistemic uncertainty is missing from standard modes of machine learning developed around statistical risk minimization and that it needs to be included when considering safety. 

We discuss a few strategies for increasing safety in machine learning that are not a comprehensive list and are far from fully developed.  This paper can be seen as laying the foundations for a research agenda motivated by safety within which further strategies can be developed and existing strategies can be fleshed out.  In some respects, the research community has taken risk minimization close to the limits of what is achievable.  Safety, especially epistemic uncertainty minimization, represents a direction that offers new and exciting problems to pursue, many of which are being pursued already.  As it is said in the Sanskrit literature, \emph{ahi\d{m}s\={a} paramo dharma\d{h}} (non-harm is the ultimate direction).  Moreover, not only is non-harm the first ethical duty, many of the safety issues for machine learning we have discussed in this paper are starting to enter legal obligations as well. For example, the European Union has recently adopted a set of comprehensive regulations for data protection, which include prohibiting algorithms that make any "decision based solely on automated processing, including profiling" and significantly affect a data subject or produce legal effects concerning him/her. This regulation which will take effect in 2018 is anticipated to restrict a wide range of machine learning algorithms currently used in, e.g., recommendation systems, credit and insurance risk assessments, and social networks\cite{GoodmanF2016}.

We present example applications where machine learning algorithms are increasingly used and discuss the aspects of epistemic uncertainty, harmful outcomes, and potential strategies for achieving safety for each application. In some applications such as cyber-physical systems and decision sciences, machine learning algorithms are used to support control and decision making in safety-critical settings with considerable costs and direct harmful impact on people's lives, such as injury or loss of life. In other applications, machine learning based predictions are only used in less critical settings for automated informational products. Applications with higher costs of unwanted outcomes tend to be also those with higher uncertainty and the ones with less severe outcomes are the ones with smaller uncertainty. 



\section{Acknowledgements}
No competing financial interests exist.



\bibliographystyle{IEEEtran}
\bibliography{MLsafety}
%



\end{document}